# RARE B AND C DECAYS, AND THE CKM MATRIX

Jon J. Thaler
University of Illinois
Urbana, IL  61801  USA
jjt@uiuc.edu

hep-ph/9507412   25 Jul 1995

I report on developments in the experimental and phenomenological understanding of the rare decays of mesons containing b and c quarks, especially as they pertain to the understanding of the CKM matrix and the testing of the standard model. Some related measurements are also discussed.

## Introduction

There have, historically, been two principal approaches to testing theories of particle physics. One has been the use of novel accelerators to study previously inaccessible or unknown phenomena. The other has been the increasingly precise measurement of known processes or particles. The recent discovery of the top quark[1] is an example of the former, while the precise measurement of $Z^o$ decays[2] is an example of the latter. The LHC will be the next novel accelerator to be built, so the next decade's progress will rely on our ability to improve experimental and theoretical precision. Fortunately, there is much activity in this direction, and I am confident that progress will continue to be made.

In this talk, I will discuss the past year's work on rare decays of B and D mesons, which has tended to focus on two large issues. The first is the improvement in the measurement of the CKM matrix elements. The origin of quark mixing remains obscure, and precise measurement may be required to reveal a pattern. The second issue is the search for phenomena which do not fit within the standard model. The discovery of CP violation is an archetypal example of such a phenomenon (in 1964, $K_L \to \pi\pi$ was a rare decay). By organizing my talk around these two issues, I will, by necessity, mention a few topics which are somewhat outside the scope implied by the title. I will also not mention a few topics which I ought to (but my talk is not allowed to run overtime). I will leave most of CP violation to Kacper Zalewski[3].

## What's new

I list here the new developments that I will discuss:

E791 (FNAL): New limits on FCNC in D decays.[4]
E687 (FNAL): New limits on CP violation in D decays.[5]
    New measurements of Cabibbo suppressed D decays.[6]
CDF (FNAL): New limits on FCNC in B decays.[7]
ALEPH (LEP): Rare hadronic B decays.[8]
DELPHI (LEP): Rare hadronic B decays.[9]
CLEO (Cornell): New limits on CP violation in D decay.[10]
    New limits on radiative D decay.[11]
    Observation of $B \to \pi \ell \nu$.[12]
    Improved measurements of rare hadronic B decays.[13]
    New measurement of $D \to \pi e \nu$[14]
BES (Beijing): Measurement of $D_S \to \mu\nu$.[15]
Phenomenology: DCSD and mixing might interfere in the D system.[16]
    Inclusive $b \to c \ell \nu$ can be used for $V_{cb}$.[17]

## CKM Matrix

Here is the CKM matrix, as given by the Particle Data Group[18], showing Wolfenstein's parameterization[19]:

$$V = \begin{pmatrix} V_{ud} & V_{us} & V_{ub} \\ V_{cd} & V_{cs} & V_{cb} \\ V_{td} & V_{ts} & V_{tb} \end{pmatrix} \cong \begin{pmatrix} 1 - \frac{\lambda^2}{2} & \lambda & A\lambda^3(\rho - i\eta) \\ -\lambda & 1 - \frac{\lambda^2}{2} & A\lambda^2 \\ A\lambda^3(1 - \rho - i\eta) & -A\lambda^2 & 1 \end{pmatrix}$$

$$= \begin{pmatrix} 0.94747 - 0.94759 & 0.218 - 0.224 & 0.002 - 0.005 \\ 0.218 - 0.224 & 0.9738 - 0.9752 & 0.032 - 0.048 \\ 0.004 - 0.015 & 0.030 - 0.048 & 0.9988 - 0.9995 \end{pmatrix}$$

These are 90% confidence level intervals. Some are known very well and some hardly at all. Also, some of the values given are calculated, in the absence of direct experimental data, by assuming 3-generation unitarity. We desire not only to improve the accuracy, but also to reduce the reliance on standard model assumptions. The parameterization itself needs to be tested. For example, to what accuracy does $|V_{cb}| = |V_{ts}|$? Some issues related to the accuracy of the Wolfenstein parameterization are discussed by Buras, *et al.*[20] Figure 1 shows the usual graphical representation of the relation between $\rho$, $\eta$, and the CKM matrix elements. The fact that $\rho$, $\eta$, and A are all O(1) justifies the form chosen.

There has been some new experimental data in the past year, but the biggest changes have come from improved theoretical understanding. $V_{cb}$ is now known about twice as well as before, and there are the first beginnings of measurements of $V_{ts}$ and $V_{tb}$.

I will have little to say about the "Cabibbo" (2x2) part of the matrix. It is measured to be:

| | |
|---|---|
| $V_{ud}$ = 0.9744 ± 0.0010 | Beta decay, |
| $V_{us}$ = 0.2205 ± 0.0018 | $K \to \pi e \nu$, |
| $V_{cs}$ = 1.02 ± 0.18 | $D \to K e \nu$, |
| $V_{cd}$ = 0.204 ± 0.017 | Charm production by neutrinos. |

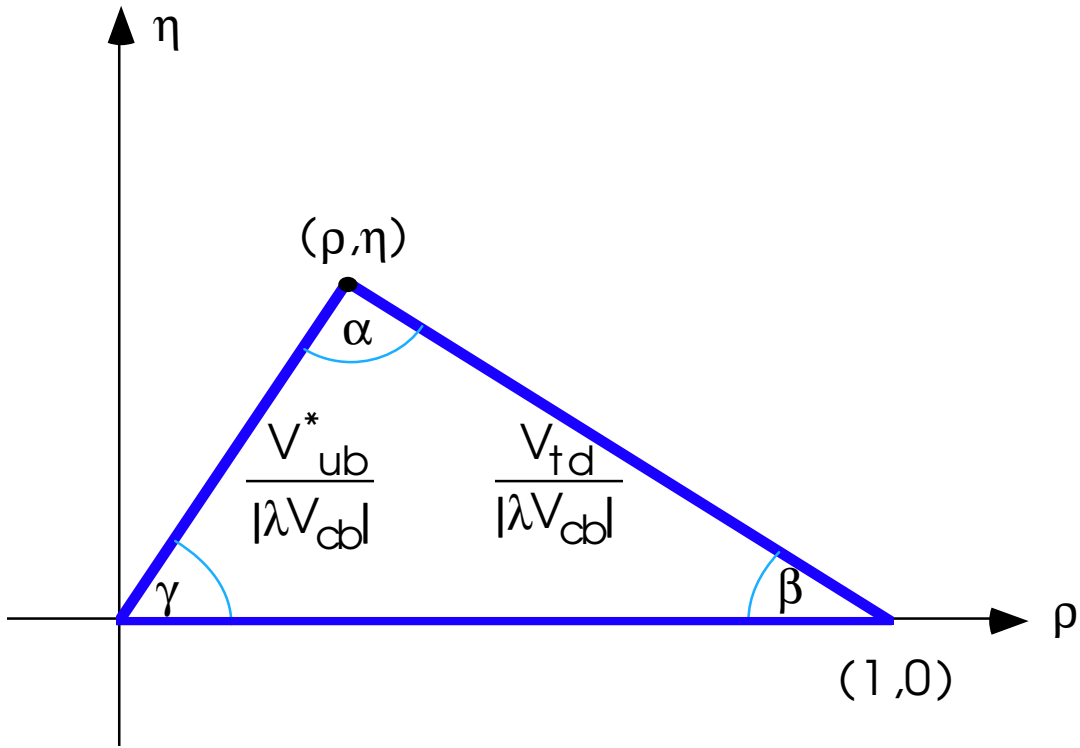

Figure 1: One unitarity triangle that can be formed with the CKM matrix.

The measurement of $V_{cd}$ is not improved by recent CLEO results, $B(D\to\pi e\nu)/(D\to K e\nu)$ = 0.103 ± 0.039 ± 0.013, because $D\to K e\nu$ background contaminates the signal. This situation is not likely to change significantly until the CLEO-III upgrade,[21] which will have improved particle identification, is complete. It is sobering to note that the accuracy of the 2x2 matrix is not sufficient to infer the presence of a third generation solely from unitarity considerations. Unless there is something vastly wrong with $V_{tb}$, it seems unlikely that a fourth generation will be discovered by studying the CKM matrix.

At the present time, rare B decays are comparable with K meson CP violation in their ability to determine the Wolfenstein parameters $\rho$ and $\eta$.[22] Figure 2 shows the current situation. Modest improvement in the B system will show that $\eta \_ 0$ without relying on K mesons. A comparison of K and B decays will test consistency of the model, especially when $K^+ \to \pi^+ \nu\nu$ is measured. For the next few years more progress will be in the B system, at least partly because more effort is going into that work.

Various B processes allow direct measurements of four of the five outer CKM matrix elements. I will discuss each of them here. Glasgow results are taken from Ali and London.[23]

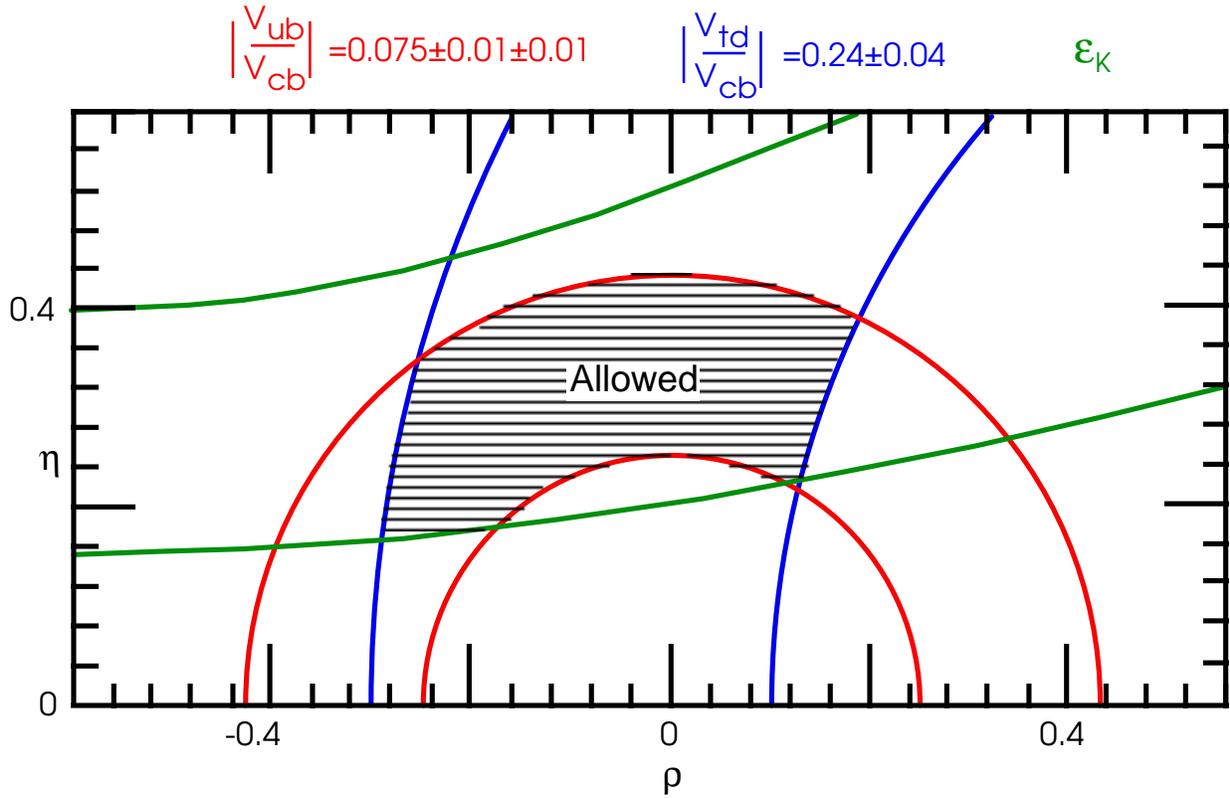

Figure 2: Allowed region of the ρ-η plane.

## Vub

Processes such as $B \to \pi \ell \nu, \rho \ell \nu, \pi\pi, \rho\pi$, which have only light quarks in the final state, are sensitive to $V_{ub}$. The best measurements have come from inclusive $b \to u \ell \nu$ decays[24] (see figure 3). The useful end point of the lepton spectrum is only a small part of the phase space, and there are severe theoretical uncertainties in the analysis. $|V_{ub} / V_{cb}| = 0.075 \pm 0.02 \Rightarrow V_{ub} = (3.1 \pm 0.9) \cdot 10^{-3}$. This spring, CLEO announced the observation of the exclusive $\pi \ell \nu$ mode,[12] with a branching ratio of $(1.70 \pm 0.51 \pm 0.31 \pm 0.27) \cdot 10^{-4}$. They also obtained $B(\rho \ell \nu)/B(\pi \ell \nu) < 3.4$. Figure 4 shows the $\pi \ell \nu$ result.

CLEO determines the neutrino momentum by fully reconstructing the entire event. The large data sets which this method requires are only now becoming available. The result is somewhat model dependent because of acceptance limitations. I quote the branching ratio which uses the WSB model, since ISGW is incompatible with the $\rho \ell \nu$ upper limit. Theory predicts[25] $B(\pi \ell \nu) = (5 - 12)|V_{ub}|^2$. So, $V_{ub} \sim (4.5 \pm 2.3) \cdot 10^{-3}$, with comparable experimental and theoretical uncertainties. This is not yet a competitive method, but its usefulness will increase with detector upgrades (background reduction) and more theoretical work.

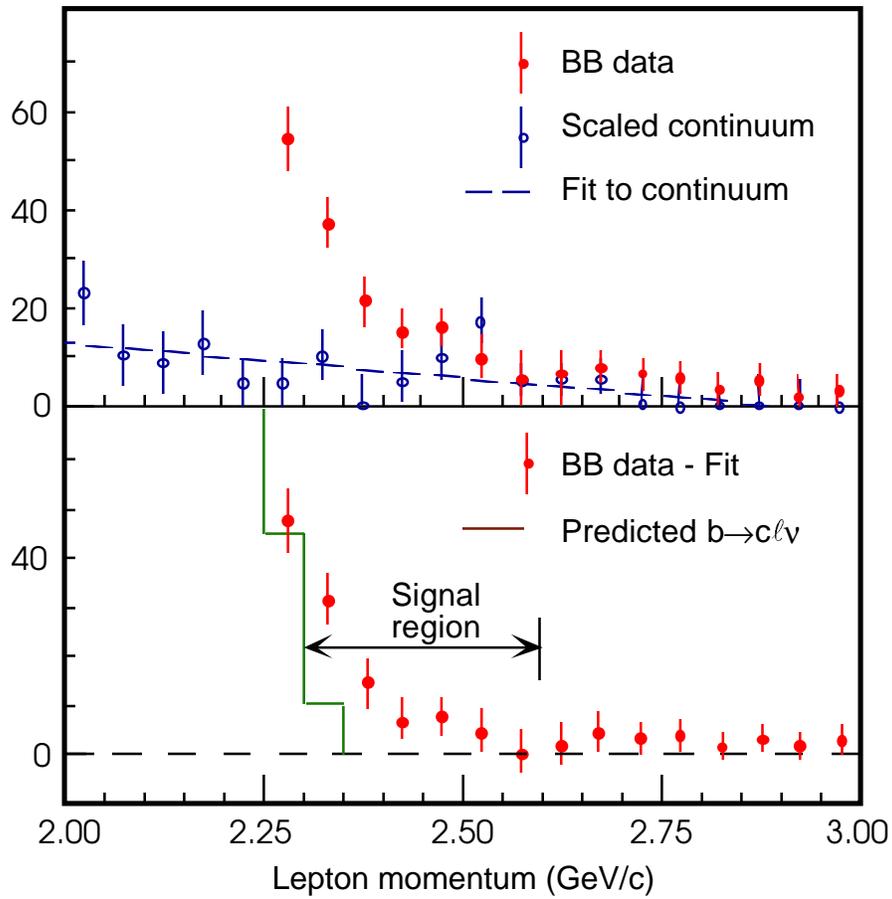

Figure 3: Inclusive measurement of $b \to u\ell\nu$.

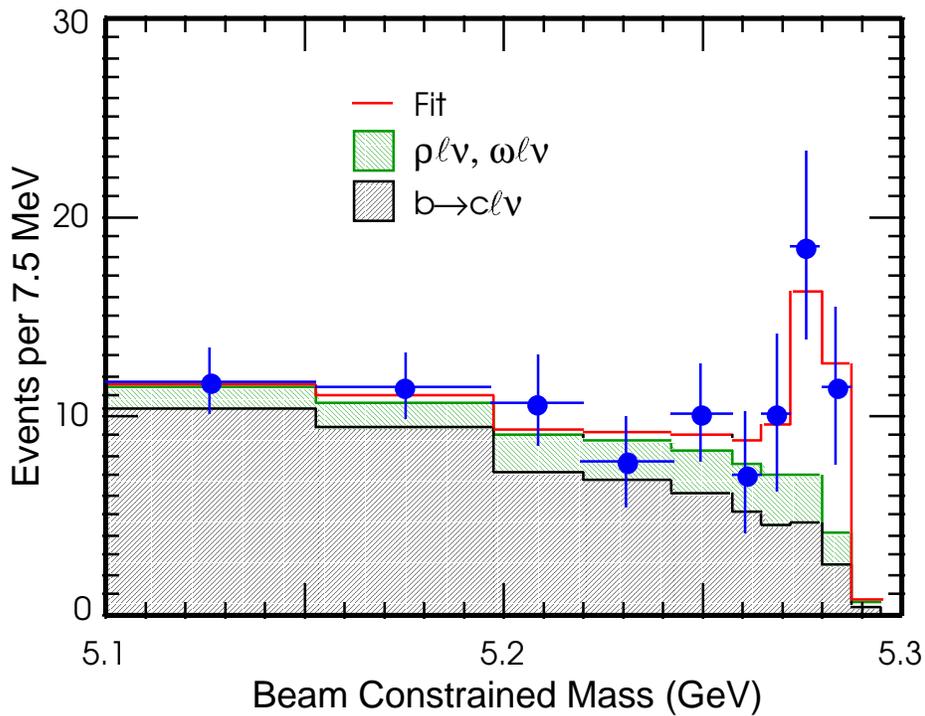

Figure 4: Beam constrained $\pi\ell\nu$ mass in $B\bar{B}$ events.

## $V_{cb}$

$V_{cb}$ controls most B decays, such as $B \to D\pi, D\ell\nu,$ and $D^*\ell\nu$, so, strictly speaking, this quantity falls outside a rare decay talk. I will mention it briefly. There are no new experimental results, but there is an important theoretical development. A typical old result[26] is CLEO's measurement $\Gamma(B \to D^*\ell\nu) = (29.9 \pm 1.9 \pm 2.7 \pm 2.0)\text{ns}^{-1}$. This yields $F(1)V_{cb}$ = 0.0351 ± 0.0019 ± 0.0018 ± 0.0008, and HQET tells us $F(1) = 0.91 \pm 0.04$[27]. At Glasgow we had $V_{cb}$ = 0.039 ± 0.006. Shifman and Ultrasev[17] have observed that inclusive $c\ell\nu$ also provides useful information about $V_{cb}$. Neubert,[27] and Ball, Benke, and Braun[28] have used this to obtain $V_{cb}$ = 0.041±0.003, a significant improvement.

## $V_{tb}$

There has until now been no data with a direct bearing on $V_{tb}$. The PDG number is obtained from unitarity constraints. CDF has now measured[29] the probabilities for $t\bar{t}$ events to have 0, 1, or 2 b-quark jets. They obtain $\frac{BF(t \to Wb)}{BF(t \to Wq)} = 0.87^{+0.13+0.13}_{-0.30-0.11}$, which implies $|V_{tb}| > 0.016$ if one assumes that there are no unknown decay channels. The limit is very weak, because $V_{tb}$ is being played against $V_{ts}$ and $V_{td}$. A significantly better limit will require a measurement of $q\bar{q} \to t\bar{b}$ after the Main Injector is operational at Fermilab. Stelzer and Willenbrock estimate[30] that 10% accuracy can be obtained with 3 fb$^{-1}$ of integrated luminosity.

## $V_{td}$

CKM matrix elements which do not have a "b" in the subscript may nevertheless be measured in B decays which depend on loops. For example, if $V_{tb}$ = 1 then $B\bar{B}$ mixing tells us[20,22]

$$|V_{td}| = 8.7 \cdot 10^{-3} \left[\frac{200\text{MeV}}{\sqrt{B_B}f_B}\right]\left[\frac{170\text{GeV}}{m_t}\right]^{0.76}\left[\frac{x_d}{0.72}\right]^{0.5}\left[\frac{1.5\text{ps}}{\tau_B}\right]$$

$$= (9.9 \pm 1.8) \cdot 10^{-3}$$

For this purpose $m_t$, $x_d$, and $\tau_B$ are all known quite well, 5.3%, 5.7%, and 3.8% respectively. $\sqrt{B_B}$ is known to 5-10% as well, so $f_{B_d}$ dominates the uncertainty. I will discuss the determination of $f_B$ and $f_D$ below. A hundred thousand t-quark events would allow LHC to measure $V_{td}$ directly. This is a difficult measurement.

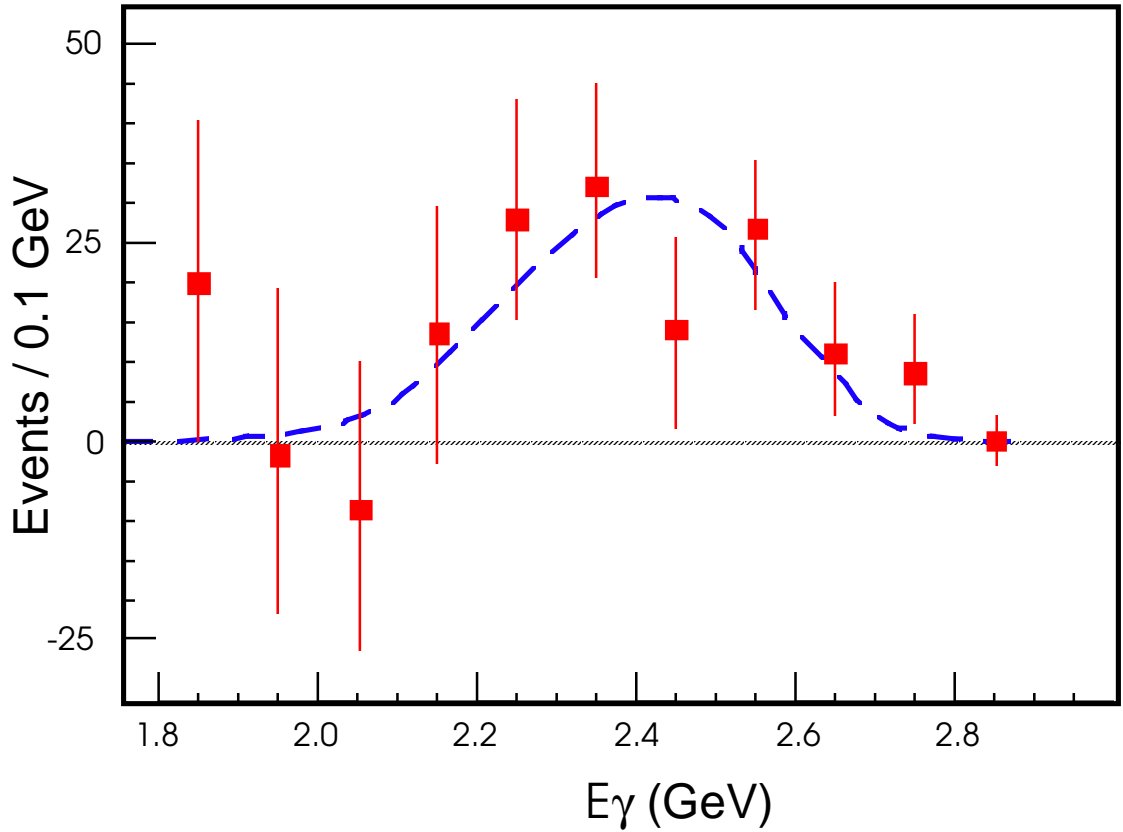

Figure 5: Inclusive b→sγ photon energy spectrum observed by CLEO. The curve is the theoretical prediction.

### $V_{ts}$

The CLEO observation[31] of the electromagnetic penguin, b→sγ, (see figure 5) has been used by Griffin, Masip, and McGuigan[32] to extract $V_{ts}$ = 0.026 ± 0.006 ± 0.011. The amplitude for this process is proportional to $V_{tb}V_{ts}^*$, so they assume that $V_{tb}$ = 1 (see the discussion above). They also must determine the form factors by extrapolating from $D \to K^* \ell \nu$. $B_s \bar{B}_s$ mixing is a more promising long term method, but at this time there is only a limit, $x_s \geq 9$.[33]

### $f_{B_d}$

The interpretation of $B\bar{B}$ mixing data is limited by the knowledge of the heavy-light decay constant $f_{B_d}$. There is no direct experimental measurement of this number; only upper limits exist for the most promising decay mode, $B \to \tau \nu$.

| Experiment | Limit on B→τν |
|---|---|
| ALEPH[34] | 1.8•10-3 |
| ARGUS[35] | 10.4•10-3 |
| CLEO[36] | 2.2•10-3 |

These results are background limited. CLEO may have improved sensitivity with its new silicon vertex detector, which will allow separation of τ from charm and other backgrounds. It remains to be seen whether CLEO can achieve the factor of 100 required to detect the anticipated $4 \cdot 10^{-5}$ branching ratio.

In the absence of a measurement, one relies on more indirect methods. Much effort has gone into calculating $f_B$ and $f_D$. Recent results are promising, but the uncertainties are still large. I take some lattice results from a review talk by S. Güsken.[37]

| Group | $f_D$ (MeV) | $f_{D_s}$ (MeV) | $f_{B_d}$ (MeV) |
|---|---|---|---|
| PSI-WUI[37] | 170 ± 30 | | 180 ± 50 |
| BLS[37] | 208 ± 35 ± 12 | | 147 ± 6 ± 23 |
| MILC[38] | 180 ± 4 ± 18 ± 16 | 194 ± 3 ± 16 ± 9 | 148 ± 4 ± 14 ± 19 |
| LANL[39] | 230 to 240 | 260 to 270 | |
| Neubert, et al.[40] | | 281 ± 44 | |

In general, $f_{Q_s}$ is about 10% larger than $f_{Q_d}$. The last result is not a lattice calculation, but the use of factorization and $B(B^o \to D^+ D_s) / B(B^o \to D^+ \pi^-) = 4.63 \pm 1.45$.

The values of $f_{D_s}$ are especially interesting, because $D_s \to \mu \nu$ has been measured.

| Group | $f_{D_s}$ (MeV) |
|---|---|
| WA75[41] | 232±45±20±48 |
| CLEO[42] | 344±37±52±42 |
| BES[15] | $430^{+150+40}_{-130-40}$ |

Only the BES result is new since last summer. They observe 3 events. CLEO has about 80 events, but the background, mostly charm semileptonic decays, is large (see figure 6). Improved vertex reconstruction will reduce the signal to background ratio.

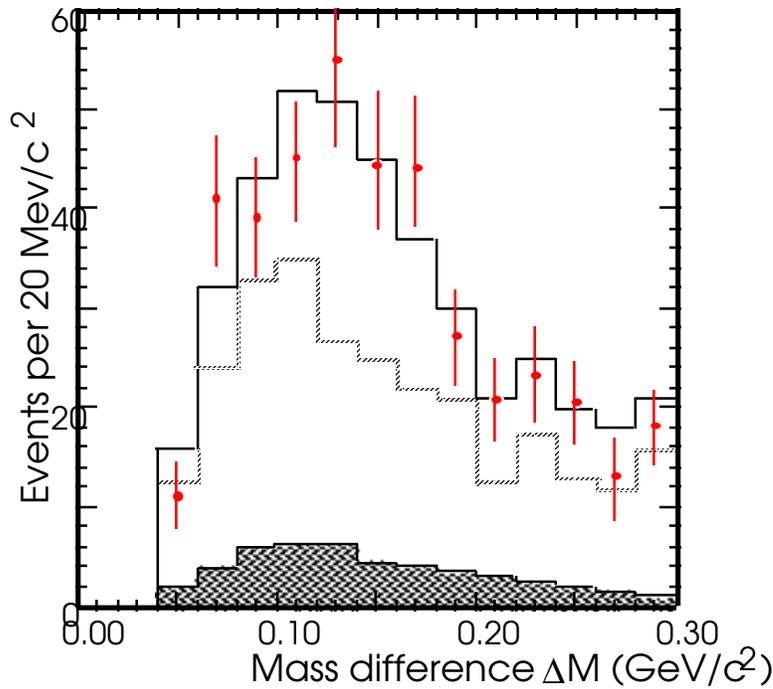

Figure 6: CLEO data for $D_s^* \to D_s \gamma \to (\mu \upsilon) \gamma$  The dashed histogram is the background, measured using electrons.

**Beyond the SM:**

Searches for new phenomena, even when unsuccessful, teach us much about particle physics. Unsuccessful searches for flavor changing neutral currents (FCNC) led to the GIM mechanism. Now, the D meson system may be the best low energy place to look for nonstandard physics, because the standard model predicts unmeasurably small mixing, CP violation, and FCNC.

**D mixing & CP:**

Mixing (or DCSD) has now been seen both by CLEO[43] and E-791.[44]

$$\frac{BR\left(D^0 \to K^+ \pi^-\right)}{BR\left(D^0 \to K^- \pi^+\right)} \begin{array}{l} = (7.7 \pm 2.5 \pm 2.5) \cdot 10^{-3} \text{ x} \\ = (2.9 \pm 1.0 \pm 1.0) \tan^4 \theta_c \quad \text{(CLEO)} \end{array}$$

$$\frac{BR\left(D^+ \to K^+ \pi^- \pi^+\right)}{BR\left(D^+ \to K^- \pi^+ \pi^+\right)} \begin{array}{l} = (10.3 \pm 2.4 \pm 1.3) \cdot 10^{-3} \\ = (3.9 \pm 0.9 \pm 0.5) \tan^4 \theta_c \quad \text{(E 791)} \end{array}$$

CLEO does not have lifetime information. That would distinguish mixing from doubly Cabibbo suppressed decays (DCSD). The best mixing limit remains that from Fermilab E615[45] using semileptonic decays: $r_{mix} < 5.6 \cdot 10^{-3}$. Other results[46] assume no interference between DCSD and mixing, which is probably incorrect.[47,48] There remains some controversy about whether the standard model predicts very small $r_{mix}$ ($^2 10^{-8}$) or quite large (~$10^{-3}$). The issue is the size of long distance corrections.[48,49] If the prediction is small, then mixing could be a signal for new physics. A fourth generation, two Higgs doublets, or leptoquarks can all cause significant mixing.[48]

E-687 and CLEO have new limits on CP violation in D decays.[50] These are still not very stringent: $A_{CP} < 5-10\%$ in several decay modes such as $D^0 \to K^+K^-$.

**B mixing and CP:**

Asymmetries due to the CP impurity of the mass states, as in the K system, are expected to be very small, because $\Delta\Gamma/\Delta M \ll 1$. For example, the standard model predicts semileptonic charge asymmetries $O(10^{-3})$ for $B_d$ and $O(10^{-4})$ for $B_s$. Direct CP violation in decay rates requires the interference between two processes leading to the same final state In order to perform consistency tests, one wants to look at various channels which measure different angles of the unitary triangle, One way to have two interfering processes is to look at states (*e.g.*, CP eigenstates) which are equally accessible from $B^0$ and $\bar{B}^0$. In that case mixing may produce interference. One difficulty is identifying the initial state. Another difficulty is that the time integrated asymmetry is zero. Asymmetric B factories will attack this problem by looking at the time dependence and tagging one B using the decays of the other.

Comparison of $B \to f$ with $\bar{B} \to \bar{f}$ does not require a time measurement, and it can be used with charged B's. Interference occurs when two different diagrams contribute to the process (*e.g.*, tree and penguin for $K^-\rho^0$). One is not restricted to $B^0$ decays. $B^\pm$ will do as well. The trick is to find decay modes which have both a significant asymmetry and a large enough branching ratio to be observable. An asymmetry also requires both weak and strong phases for the two diagrams.

New results on rare hadronic B decays, such as $B^0 \to \pi^+\pi^-$ and $K^+\pi^-$, have implications for future CP violation searches. $B^0 \to \pi^+\pi^-$ is dominated by the b→u spectator diagram. It could be used to measure $V_{ub}$ and ~~CP~~. $B^0 \to K^+\pi^-$ is primarily a hadronic penguin. The recent discussion of extracting phase angles from time integrated rates[51] is of obvious interest to CLEO, which will have high luminosity (comparable to the B factories), but equal beam energies. The idea is that ~~CP~~ could be observed in $B^\pm$ as well as $B^0$ decays, because direct ~~CP~~ is expected to be significant. An example is: $B^\pm \to K^\pm\pi^0$, $B^\pm \to \pi^\pm\pi^0$, $B^\pm \to K^0\pi^\pm$. CP violation would manifest itself as a charge asymmetry in the decay rates (see figure 7). CLEO quotes only upper limits[13] in any single exclusive channel, $(1-5)\cdot 10^{-5}$ in $\pi^+\pi^-$, $K^+\pi^-$, $\pi^+\pi^0$, $K^+\pi^0$, $\pi^0\pi^0$, $K^0\pi^+$, and $K^0\pi^0$. However, there is an unambiguous signal in the sum $\pi^+\pi^- + K^+\pi^-$. $N_{\pi\pi+KK} = 17.2^{+5.6+2.2}_{-4.9-2.5}$ events (5.2$\sigma$ from 0, see figure 8). Particle identification limitations prevent a clean separation. $B(\pi^+\pi^-$ or $K^+\pi^-) = (1.8\pm 0.6\pm 0.2)\cdot 10^{-5}$. ALEPH has 2 $\pi^+\pi^-$ events (one event reported in '94 has been cut), while DELPHI has 3 events in different modes. They both report upper limits somewhat higher than CLEO's. New results are expected this summer. An increase of a factor of ten in integrated luminosity, so that there are about 100 events in each mode, along with an improvement in particle identification, may allow a 10 degree measurement of $\gamma$.

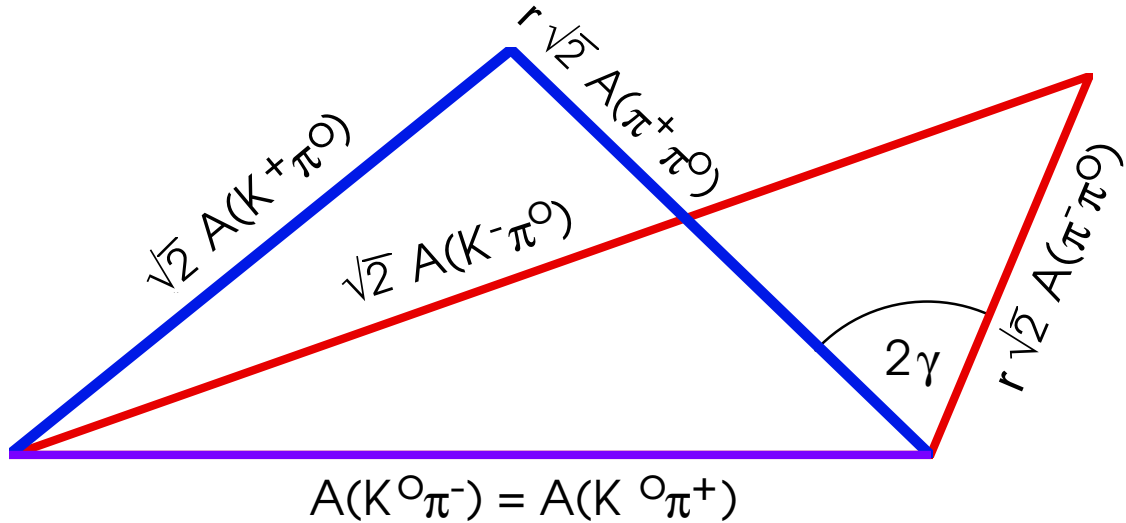

Figure 7: Amplitude triangles for three $B^{\pm}$ decay modes. CKM unitarity and SU(3) symmetry imply the triangle relationships among the decay amplitudes.

**FCNC:**

Flavor changing neutral currents ($B$ or $D \rightarrow \ell\ell$ or $X\ell\ell$) have much in common with the radiative penguin process, b,c → Xγ. Having two charged leptons in the final state, they are easier to detect. However, the decay rates are expected to be quite a bit smaller, and they suffer from significant contamination from long distance effects, especially $B \rightarrow \Psi X$. There are several new results on FCNC (see the table on the next page). All are limits, although CLEO will approach the standard model prediction for $B^0 \rightarrow K^0 e^+ e^-$ in the next year or two. Due to large long distance (vector dominance) effects in radiative D decays, only $D^+ \rightarrow \pi^+ e^+ e^-$ appears to have a new physics opportunity in the D system.

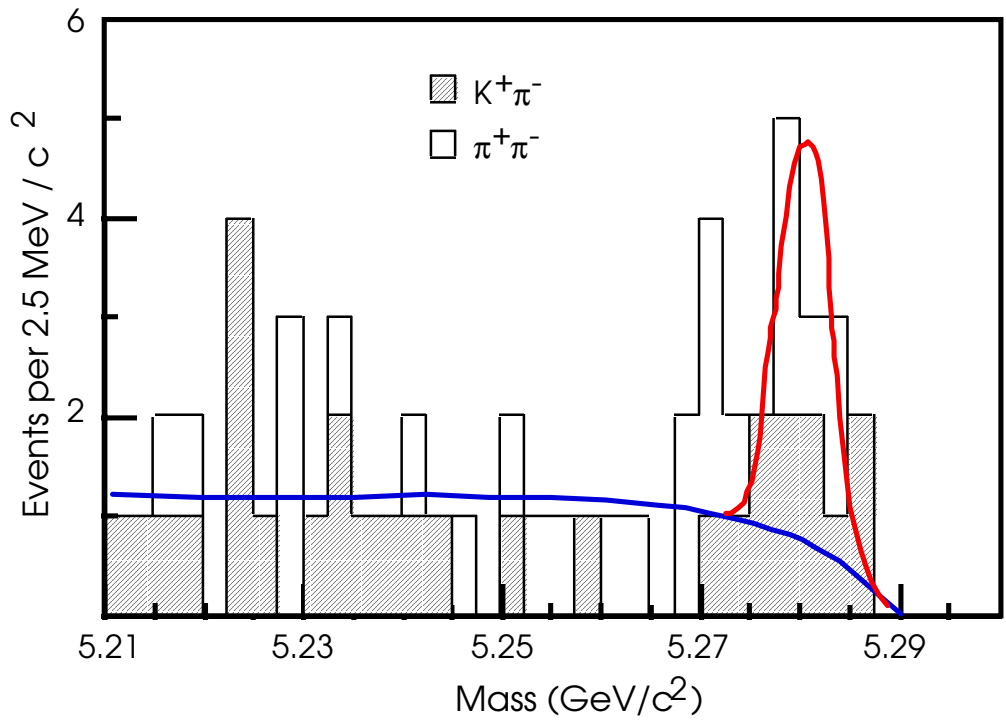

Figure 8: CLEO's evidence for B → ππ or Kπ.

| Experiment | Process | Upper limit | SM prediction[52] |
|---|---|---|---|
| CLEO[13] | $B^+ \to K^+ e^+ e^-$ | $1.2 \cdot 10^{-5}$ | $0.06 \cdot 10^{-5}$ |
| | $K^+ \mu^+ \mu^-$ | $0.9 \cdot 10^{-5}$ | $0.06 \cdot 10^{-5}$ |
| | $B^0 \to K^0 e^+ e^-$ | $1.6 \cdot 10^{-5}$ | $0.56 \cdot 10^{-5}$ |
| | $K^0 \mu^+ \mu^-$ | $3.1 \cdot 10^{-5}$ | $0.29 \cdot 10^{-5}$ |
| CLEO[11] | $D \to (\rho, \phi, K^*)\gamma$ | $(1 \text{ to } 2) \cdot 10^{-4}$ | $10^{-6}$ to $10^{-4}$ |
| CDF[7] | $B^+ \to K^+ \mu^+ \mu^-$ | $3.5 \cdot 10^{-5}$ | $0.06 \cdot 10^{-5}$ |
| | $K^{+*} \mu^+ \mu^-$ | $5.1 \cdot 10^{-5}$ | $0.23 \cdot 10^{-5}$ |
| | $B_d^0 \to \mu^+ \mu^-$ | $0.2 \cdot 10^{-5}$ | $8 \cdot 10^{-11}$ |
| | $B_s^0 \to \mu^+ \mu^-$ | $0.7 \cdot 10^{-5}$ | $2 \cdot 10^{-9}$ |
| E791[4] | $D^+ \to \pi^+ e^+ e^-$ | $6.6 \cdot 10^{-5}$ | $\sim 1 \cdot 10^{-8}$ |
| | $\pi^+ \mu^+ \mu^-$ | $1.8 \cdot 10^{-5}$ | " |

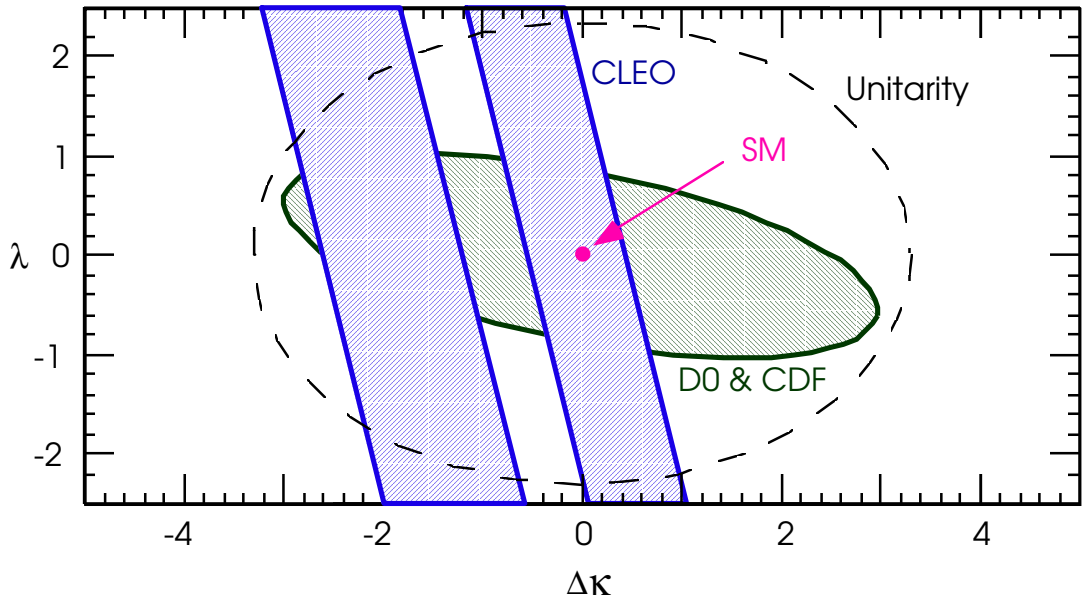

Figure 9: 90% c.l. limits on $WW\gamma$ anomalous couplings. The dashed oval is the unitarity limit for $\Lambda_W = 1.5$ TeV. Adapted from Hinchliffe.[53]

**Anomalous gauge boson couplings:**

Finally, I would like to mention a pretty result. The CLEO value[31] for $B(b\to s\gamma) = (2.32 \pm 0.57 \pm 0.35)\cdot 10^{-4}$, can be combined with CDF and D0 measurements[54] of $W\gamma$ production to constrain anomalous $WW\gamma$ couplings (see figure 9). The results are consistent with the standard model (no anomalous couplings). The limits are already better than unitarity constraints, and will continue to improve.

**Acknowledgments**

I would like to thank all those people who provided me with their data, especially, Tony Liss, Tom LeCompte, Alan Litke, Wilbur Venus, Karen Lingel, Mats Selen, Milind Purohit, Jeff Appel, and Noel Stanton. I would also like to thank Scott Willenbrock and Aida El-Khadra for stimulating discussions. Finally, I thank Tom Browder and Klaus Honscheid, whose "B Mesons" review paper provided valuable guidance.

1. Observation of the Top Quark, talk at this conference by Soo-Bong Kim.
2. *Precision Tests of Electroweak Interactions*, talk at this conference by Wolfgang Hollik.
3. *Structure and Decay of Heavy Quarks*, talk at this conference by Kacper Zalewski.
4. M. V. Purohit, J. A. Appel, and Noel Stanton, private communication.
5. P. Frabetti, *et al.*, *Phys. Rev.* **D50**,2953(1994).
6. P. Frabetti, *et al.*, *Phys. Lett* . **B321**,295(1994) and *Phys Lett* **B346**,199(1995).
7. Carol Anway-Wiese, Fermilab-Conf-94/210-E.
8. A.M. Litke, SCIPP 94/35 (1994). in *Proceedings of the XXVII International Conference on High Energy Physics* (Glasgow 1994), p. 1333, and private communication.

[9] CERN-PPE/Draft2/Paper0104. Wilbur Venus, private communication.
[10] J. Bartelt, *et al.*, CLNS 95/1333.
[11] M. Selen, talk at APS Meeting, April 1995.
[12] L. Gibbons, talk at *30th Rencontres de Moriond* (1995).
[13] S. Playfer and S. Stone, HEPSY 95-01 (hep-ph 9595392), to appear in *Intl. J. Mod. Phys. Lett.*
[14] F. Butler, *et al.*, CLNS 95/1324 (1995).
[15] J.Z. Dai, *et al.*, SLAC-PUB-95-6746 (1995), to appear in *Phys. Rev. Lett.*
[16] T. Liu, in *Proceedings of the CHARM2000 Workshop*, 375(1995),
G. Blaylock, A. Seiden, and Y. Nir, SCIPP-95-16 (hep-ph 9504306).
[17] M. Shifman, in *Proceedings of the XXVII International Conference on High Energy Physics* (Glasgow 1994), p, 1125,
M. Shifman and N. G. Ultrasev, TPI-MINN-94/41-T (hep-ph 9412398).
[18] Particle Data Group, *Phys. Rev.* **D50**, 1175 (1994)..
[19] L. Wolfenstein, Phys. Rev. Lett **51**, 1945 (1983).
[20] A. J. Buras, M. E. Lautenbacher, and G. Ostermaier, *Phys. Rev.* **D50**, 3433 (1994).
[21] K. Berkelman, CLNS 93/1265 (1993).
[22] A. J. Buras, MPI-PhT/95-17, to appear in *Acta Physica Polonica*.
[23] A. Ali and D. London, in *Proceedings of the XXVII International Conference on High Energy Physics* (Glasgow 1994), 1133 (1995),
A. Ali and D. London, CERN-TH.7398/94.
[24] H. Albrecht, *et al.*, *Phys. Lett.* **B234**, 409 (1990),
H. Albrecht, *et al.*, *Phys. Lett.* **B255**, 297 (1991),
R. Fulton, *et al.*, *Phys. Rev. Lett.* **64**, 16 (1990),
J. Bartelt, *et al.*, *Phys. Rev. Lett.* **71**, 4111 (1993).
[25] M. Wirbel, B. Stech, and M. Bauer, *Z. Phys.* **C29**, 637 (1985).
J. G. Körner and G.A. Schuler, *Z. Phys.* **C38**, 511 (1988),
P. Ball, *Phys. Rev.* **D48**, 3190 (1994),
S. Narison, CERN-TH.7237/94 (1994),
N. Isgur and D. Scora, CEBAF-TH-94-14 (1994),
R. Faustov, V.O. Galkin, and A. Yu. Mishurov, (hep-ph 9505321).
[26] B. Barish, *et al.*, *Phys. Rev.* **D51**, 1014(1995),
M. Athenas, *et al.*, *Phys. Rev. Lett.* **73**, 3503(1994),
I.J. Scott, in *Proceedings of the XXVII International Conference on High Energy Physics* (Glasgow 1994), 1121(1995),
H. Albrecht, *et al.*, *Z. Phys.* **C57**, 533(1993),
H. Albrecht, *et al.*, *Phys. Lett.* **B318**, 397(1993).
[27] M. Neubert, in *Proceedings of the XXVII International Conference on High Energy Physics* (Glasgow 1994), 1129 (1995),
M. Neubert, CERN-TH/95-107 (hep-ph 9505238),
[28] P. Ball, M. Benke, and V.M. Braun, CERN-TH/95-65 (hep-ph 9503492),
also see T.E. Browder and K. Honscheid, OHSTPY-HEP-E-95-010,
to appear in *Progress in Nuclear and Particle Physics* **35**.
[29] T. J. LeCompte and R. M. Roser, CDF/ANAL/TOP/CDFR/3056 (1995),
and private communication.
[30] T. Stelzer and S. Willenbrock, ILL-(TH)-95-30, (hep-ph 9505433).
[31] R. Ammar, *et al.*, *Phys. Rev. Lett.* **71**, 674(1993),
M.S. Alam, *et al., Phys. Rev. Lett.* **74**, 2885(1995).
[32] P.A. Griffin, M. Masip, and M. McGuigan, *Phys. Rev. D* **50**, 5751 (1994).
[33] Yi-Bin Pan, in *Proceedings of the XXVII International Conference on High Energy Physics* (Glasgow 1994), p. 753,
S. Komamiya, *ibid.*, p. 757.
[34] D. Buskulic, et al., *Phys. Lett.* **B343**, 444 (1994).
[35] H. Albrecht, *et al.*, DESY 94-246.
[36] J. Alexander, et al., CLEO-CONF-94-5, to appear in *Phys Rev Lett*.
[37] S. Güsken, WUB 95-08, talk at *Workshop on Heavy Quarks*.
[38] C. Bernard, et al., FSU-SCRI-95C-28, talk at LISHEP95 (hep-ph 9503336).
[39] T, Bhattacharya and R. Gupta, hep-lat 9501016.
[40] M. Neubert, V, Rieckert, B. Stech, Q.P. Xu, in *Heavy Flavors* (A. Buras and M. Lindner, eds.), p. 286 (1992).

41. S. Aoki, *et al.*, *Prog. Theor. Phys.* **89**, 131 (1993).
42. D. Acosta, *et al.*, *Phys. Rev.* **D49**, 5690 (1994).
43. D. Cinabro, *et al.*, *Phys. Rev. Lett.* **72**, 1406(1994),
44. M. Purohit and J. Weiner, FERMILAB-CONF-94/408-E, to appear in the *Proceedings of the Eight Meeting of the DPF of the APS* (DPF '94).
45. W.C. Louis, *et al.*, *Phys. Rev. Lett.* **56**, 1027(1986).
46. M.V. Purohit, et al., FERMILAB-CONF-94/186-E.
47. T. Liu, in the *Proceedings of the CHARM2000 Workshop*, 375(1995),
    G. Blaylock, A. Seiden, and Y. Nir, SCIPP-95-16 (hep-ph 9504306).
48. J.L. Hewett, SLAC-PUB-95-6821 (hep-ph 9505246).
49. T.A. Kaeding, LBL-37224 (hep-ph 9505393),
    L. Wolfenstein, Phys. Lett. **B164**, 170(1985),
    L. Wolfenstein, CMU-HEP95-04 (hep-ph 9505285).
50. P. Frabetti, *et al.*, *Phys. Rev.* **D50**,2953(1994),
    J. Bartelt, *et al.*, CLNS 95/1333.
51. M. Gronau, J.l. Rosner, and D. London, *Phys. Rev. Lett.* **73**, 21(1994),
    N. G. Deshpande and X-G. He, *Phys. Rev. Lett.* **74**, 26(1995),
    M. Gronau, *et al.*, TECHNION-PH-95-10, (hep-ph 9504327),
    O.F. Hernandez, UdeM-GPP-TH-95-20, (hep-ph 9502335).
52. A. Ali and T. Mannel, *Phys. Lett.* **B264**, 446 (1991),
    A. Ali, C. Greub, and T. Mannel, DESY-93-016 (1993),
    G. Burdman, E. Golowich, J. L. Hewett, and S. Pakvasa, SLAC-PUB-6692 (hep-ph 9502329),
    A. J. Schwartz, *Mod. Phys. Lett.* **A8**, 967 (1993).
53. I. Hinchliffe, LBL-37014 (hep-ph 9504206), talk at *International Symposium on Vector Boson Self Interactions* (1995).
54. F. Abe, et al., *Phys. Rev. Lett.* **74**, 1936 (1995),
    S. Abachi, FERMILAB-PUB-95-101-E (hep-ex 9505007).